\documentclass[pre,twocolumn,groupedaddress]{revtex4-1}

\usepackage{graphicx,bbm,braket,color,lipsum,amsmath}
\begin{document}

%\preprint{}

\title{Modeling of nonequilibrium surface growth by a limited mobility model with distributed diffusion length}


\author{Thomas Martynec}
\email[]{martynec@tu-berlin.de}
\author{Sabine~H.~L.~Klapp}
\email[]{klapp@physik.tu-berlin.de}
%\thanks{}
\affiliation{Institut f\"ur Theoretische Physik, Technische Universit\"at Berlin, Hardenbergstr. 36, 10623 Berlin, Germany}


\date{\today}

\begin{abstract}

Kinetic Monte-Carlo (KMC) simulations are a well-established numerical tool to investigate the time-dependent surface morphology in molecular beam epitaxy (MBE) experiments. In parallel, simplified approaches such as limited mobility (LM) models characterized by a fixed diffusion length have been studied. Here, we investigate an extended LM model to gain deeper insight into the role of diffusional processes concerning the growth morphology. Our model is based on the stochastic transition rules of the Das Sarma-Tamborena (DT) model, but differs from the latter via a variable diffusion length. A first guess for this length can be extracted from the saturation value of the mean-squared displacement calculated from short KMC simulations. Comparing the resulting surface morphologies in the sub- and multilayer growth regime to those obtained from KMC simulations, we find deviations which can be cured by adding fluctuations to the diffusion length. This mimics the stochastic nature of particle diffusion on a substrate, an aspect which is usually neglected in LM models. We propose to add fluctuations to the diffusion length by choosing this quantity for each adsorbed particle from a Gaussian distribution, where the variance of the distribution serves as a fitting parameter. We show that the diffusional fluctuations have a huge impact on cluster properties during submonolayer growth as well as on the surface profile in the high coverage regime. The analysis of the surface morphologies on one- and two-dimensional substrates during sub- and multilayer growth shows that the LM model can produce structures that are indistinguishable to the ones from KMC simulations at arbitrary growth conditions.

\end{abstract}

%\keywords{}

\maketitle

\section{Introduction}

Nonequilibrium surface growth by means of molecular beam epitaxy (MBE) is one of the most widely used techniques to fabricate thin film devices for various technological applications \cite{Yao2016, Wofford2017, Jin2016, Tobin1990}. Since the growth conditions can be precisely controlled, MBE also serves as an exemplary experimental setup to study fundamental aspects of nonequilibrium statistical mechanics \cite{Barabasi1995,Michely2003,Pimpinelli1998}.

The time-dependent morphologies in MBE evolve due to a competition between adsorption of particles on the system's surface, on the one hand, and diffusion processes, on the other hand. Particles like atoms, organic molecules or colloids get adsorbed on a flat and defect-free substrate (ideal growth conditions) at rate $F$, which is typically given in deposited monolayers (ML) per second (ML/s) \cite{Evans2006,Barabasi1995,Pimpinelli1998,Michely2003}. The adsorption is followed by thermally activated (Arrhenius-type) diffusion processes with energy-dependent rates $D(T) = \nu_{0} \text{exp} (-E_{A}/k_{B}T)$, where $\nu_{0} = 2k_{B}T/h$ is the attempt frequency (with $k_{B}$ being the Boltzmann constant, $h$ the Planck constant and $T$ the substrate temperature) and $E_{A}$ the activation energy that consists of different energetic contributions. ''Free'' particles, i.e., particles without in-plane bonds towards particles on neighboring sites, diffuse laterally on the substrate at rate $D^{0}(T)$ until they participate in a cluster formation event (nucleation), or attach to an existing cluster. Depending on the details of the system (i.e., the type of deposited particles and substrate material) and the growth conditions, that is, the temperature $T$ and adsorption rate $F$, the surface evolves either smooth or rough. Commonly, the growth conditions are expressed by the growth parameter 

\begin{equation}
R = R(T,F) = D^{0}(T) / F.
\end{equation}

There are various conceptually different simulation strategies to model the emerging morphologies in nonequilibrium surface growth, including the level-set method \cite{Vvedensky2002,Osher2001}, geometry-based approaches \cite{Evans2003}, molecular dynamics simulations \cite{Spraque1991,Gilmore1992,Spraque1993,Moseler1995,Srolovitz1996,Ferrando2000,Burgos2005,Chirita2016} and numerical solutions of stochastic equations governing the evolution of the surface height \cite{Haselwandter2005,Haselwandter2006,Haselwandter2007_1,Haselwandter2007_2,Haselwandter2008,Haselwandter2010}. One further, very popular simulation strategy is to employ lattice models that are based on activation energy-dependent hopping rates for all particles in the topmost layer. These models are often referred to as ''full diffusion'' or Arrhenius-type models \cite{Weeks1979,Maksym1988,Kotrla1996,Levi1997,Kleppmann2015_1,Martynec2018,Kleppmann2015_2,Klopotek_I,Kleppmann2016,Klopotek_II,Jana2013}. One major example is the (event-driven) kinetic Monte-Carlo (KMC) method based on the Clarke-Vvedensky bond-counting Ansatz \cite{ClarkeVvedensky98} involving diffusion to nearest-neighbor lattice sites. Concerning atomic systems, not only the experimental morphologies seen in MBE experiments can be reproduced by KMC simulations, but also fine details of the growth process at the atomistic length scale \cite{Hohage1996,Evans1995,Family1996,Evans2002,Evans2008,Kratzer2002,Godbey1994,Nie2017}. More complicated is the growth of organic systems where essentially only the morphologies can be described \cite{Krause2004,Choudhary2006,Bommel2014,Acevedo2016} due to the generally more complicated interparticle interactions involved.

Even though the KMC method can nowadays handle growth simulations with large growth parameters and many deposited layers, they still require a significant amount of computational time. This is mainly due to the computational effort required to simulate the trajectories of freely diffusing particles, without making much progress in the actual time evolution of the system. To speed up the simulations, multiscale approaches, where the fastest dynamical process involved (i.e., free lateral diffusion) is described in an averaged mean-field manner or by an appropriate diffusion equation, have been introduced and investigated in detail \cite{Schoell2003,Smereka2005,Opplestrup2006,Chou2006,Tokar2008,Tokar2009,Tsalikis2013}.

An alternative class of systems to model nonequilibrium surface growth are discrete lattice growth models which are known as limited mobility (LM) models \cite{Family1990,Wolf1990,Villain1991,DasSarma1991,DasSarma1992}. Due to their simplicity, these models are especially suitable to investigate scaling properties, to study kinetic surface roughening and morphological properties as well as to investigate details like crossover and long-lived transient effects in nonequilibrium surface growth \cite{Chatraphorn2001,Chatraphorn2002,DasSarma2002, DasSarma2000}. In LM models, the process rates that are used in KMC simulations are replaced by a certain set of stochastic rules for particle movements that depend on the local environment of the position of particle adsorption. Importantly, the deposited particles only perform one single movement that depends on the specific rules of the underlying LM model. Well-known examples of LM models with surface diffusion include the Family (F) model \cite{Family1990}, the Wolf-Villain (WV) model \cite{Wolf1990, Villain1991} and the model of Das Sarma and Tamborenea (DT) \cite{DasSarma1991,DasSarma1992}.

In the present study, we introduce an extended version of the DT model since the latter is particularly suitable to describe \textit{low} temperature MBE growth (detachment processes can be essentially neglected). In the original version of the DT model \cite{DasSarma1991}, adsorbed particles only explore the nearest neighbors of the adsorption site. This scenario corresponds to a diffusion length $l = 1$ (in units of the lattice constant). However, under realistic conditions for MBE growth, the average diffusion length of particles is usually $l > 1$, a situation that has been studied in the literature in different variants \cite{Mal2010, Mal2017, DasSarma2002,To2018,Reis2010}. Studying the case $l > 1$ generally requires the use of various fit parameters in the chosen LM model to match the results of corresponding KMC simulations \cite{Reis2010,To2018}. Here, we employ a LM model with fit parameters that are based on physical quantities only. Extending the DT model towards $l > 1$ implies that we have to find a prescription of how to choose $l$ for a given value of $R$ [see Eq. (1)]. This is the first objective of the present paper.

More specifically, we aim to choose the value of $l$ based on an appropriate quantity calculated by (short) KMC simulations in the submonolayer growth regime. In other words, we seek a mapping procedure between the two type of models. The goal is that the resulting LM model produces surface structures indistinguishable to those obtained from KMC simulations (and therefore also identical to \textit{low} temperature MBE growth) at any value of $R$ with, at the same time, highly reduced computational effort. In this way, the LM model can be used to simulate MBE at growth conditions and system sizes that are typically hard to access in KMC simulations, especially when averaging over many realizations is required. This would enable us, for example, to study the asymptotic regime of the surface growth where we expect to observe scaling behavior of the growing surface. In particular, one would like to extract the corresponding critical exponents describing the scaling of the surface roughness \cite{Family_Vicsek_1985} without being limited by finite-size effects or computational manipulations like the noise reduction technique (NRT) \cite{Kim1989,Wolf1989,Punyindu1998,Punyindu2001,Punyindu2002}.

The second main goal of this study is to investigate how the strength of fluctuations of the diffusion length in the model with limited particle mobility alters the resulting surface morphologies, as compared to the case where the diffusion length is the same for all particles that are deposited during the growth process.

The remainder of the manuscript is structured as follows. In Sec. II, the KMC model and the LM model are introduced and explained in detail. In line with other studies in this area \cite{Mal2010, Mal2017, DasSarma2002,To2018,Reis2010}, we mainly focus on the one-dimensional case, but consider two dimensional lattices as well. Following this, we establish in Sec. \ref{relation_R_l} a relation between $R$ and $l$ to connect both models. A numerical analysis and comparison of the two models in the sub- and multilayer growth regime at various growth conditions is given in Sec. IV. There, we also highlight the importance of diffusional fluctuations in the regime of large values of $l$ and investigate the general effect of a variable diffusion length on the surface morphology in the multilayer growth regime. Results of our approach in two dimensions are presented in Sec. V. Finally, we summarize and conclude in Sec. VI.

\section{Simulations details}

\subsection{System settings}

Simulations in this study were performed on one- and quadratic two-dimensional substrates ($d = 1,2$) with discrete, equidistant positions $i,j = 1,2,...,L$. The corresponding local surface heights in one-dimension are given by the integers $h_{i}$ and by $h_{ij}$ in two dimensions (i.e., $h_{i} = 0$ corresponds to an empty site).

We apply periodic boundary conditions and the solid-on-solid condition, that is, vacancies and overhanging particles are not allowed. As a consequence, the spatially averaged surface height on the one-dimensional lattice at time $t$ is given by

\begin{equation}
\braket{h(t)} = \frac{1}{L} \sum_{i=1}^{L} h_{i}(t) = Ft 
\end{equation}

where the expression on the right side corresponds to the number of deposited particles. Generalization to the two-dimensional case is straightforward. The product $Ft$ is henceforth referred to as coverage $\theta = Ft$. Therefore, time-dependent quantities can also be expressed as functions of $\theta$. Throughout this work, we characterize the growth conditions via the dimensionless, free diffusion to adsorption ratio $R(T,F)$ defined in Eq. (1).

\subsection{The kinetic Monte-Carlo model}

Within the KMC method, particles are adsorbed on randomly chosen lattice sites with an (effective) adsorption rate $F$. The adsorption process is followed by diffusion processes to nearest-neighbor lattice sites. Following the Clarke-Vvedensky bond-counting Ansatz \cite{ClarkeVvedensky98}, the hopping rates are given by 

\begin{equation}
D(T) = \nu_{0} \text{exp}(-E_{A}/kT), 
\end{equation}

with activation energy $E_{A} = E_{D} + n E_{N}$. Here, $E_{D}$ is the energy barrier for free diffusion, which we set to $E_{D} = 0.5$ eV in all KMC simulations in this study. We use this value for $E_{D}$ because it is close to the known diffusion barriers of various, intensely studied, atomic and organic systems \cite{Hohage1996, Bommel2014}, and because this choice is consistent with previous KMC studies \cite{Bommel2014, Martynec2018}. The rate for free diffusion to neighboring lattice sites is then given by 

\begin{equation}
D^{0}(T) = \nu_{0} \text{exp}(-E_{D}/kT). 
\end{equation}

The additional energy contribution $E_{N}$ stems from interactions with nearest-neighbors in lateral directions. Here, $n$ is the number of such bonds. In one dimension, this number can take the values $n = 0,1,2$, while $n = 0,1,2,3,4$ in two dimensions. We here choose a high value of $E_{N}$, that is, $E_{N} = 1.0$ eV, in order to mimic MBE growth at \textit{low} $T$. Then, already one in-plane bond is sufficient to suppress further diffusion. In other words, particles immediately immobilize once they sit on a lattice site $i$ with $n > 0$. Consequently, already dimers represent stable clusters, and the critical cluster size $i^{*}$ is one \cite{Amar_Family_95}.

A typical KMC simulation consists of a large number of iterations $p$. In each iteration step, a particle either performs a hopping process to a neighboring lattice site, or a new particle gets adsorbed. The simulation time (with $t_{0} = 0$ being the starting time) after $p$ iteration steps is updated stochastically as

\begin{equation}
t_{p+1} = t_{p} + \tau,
\end{equation}

where $\tau$ is defined as 

\begin{equation}
\tau = - \frac{\text{ln}(X)}{r_{\text{all}}}.
\end{equation}

Here, $X \in (0,1)$ is a random number chosen uniformly from the given interval, and 

\begin{equation}
r_{\text{all}} = \sum_{i=1}^{L} \left( \sum_{j = 1}^{2} D_{ij} + F \right)
\end{equation}

is the sum of rates related to all particles in the topmost layer of the discretized (one dimensional) lattice. Again, the generalization to a two-dimensional lattice is straightforward.

For simplicity, we do not consider an additional energy barrier $E_{ES}$ for inter-layer diffusion processes across step-edges, usually referred to as Ehrlich-Schwoebel barrier \cite{Ehrlich1966,Schwoebel1966,Schwoebel1969}. However, such a barrier could be included, in principle. The temperature is fixed to $T = 273$ K in all KMC simulations in this study. In order to realize different growth conditions expressed via the growth parameter $R = D^{0}(T) / F$, we use $F$ as a variable.

\begin{figure}
	\includegraphics[width=1.0\linewidth]{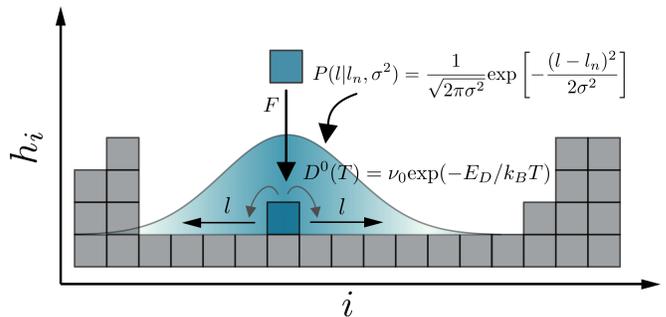}
	\caption{\label{Average cluster shape} Illustration of particle deposition and surface relaxation in the system (in one dimension) that is used to simulate low temperature MBE growth. The quantities $F$ and $D^{0}(T)$ refer to the KMC simulation, while the Gaussian distribution $P(l|l_{n}, \sigma^{2})$ for the diffusion length $l$ is characteristic for the LM model with diffusional fluctuations whose strength is controlled via the variance $\sigma^{2}$.}
\end{figure}

\subsection{The limited mobility model}

The second model we investigate falls into the class of limited mobility models. Specifically, we consider a variant of the intensively studied model by Das Sarma and Tamborenea (DT model), in which the diffusion length is restricted to one lattice constant ($l = 1$) \cite{DasSarma1991,DasSarma1992}. In contrast, here we consider the case $l \ge 1$ \cite{DasSarma2002,Reis2010,To2018} and additionally, consider $l$ as a fluctuating quantity.

To clarify our approach, we first summarize the algorithm of the original DT model on a one-dimensional lattice. In each iteration step, a particle is adsorbed at a randomly chosen lattice site $i \in [1,L]$ and sticks there permanently if it has at least one in-plane nearest-neighbor. Otherwise, the particle is allowed to hop either to the left neighboring site, $j = i-1$, or to the right neighboring site, $j = i+1$, if one of these two sites provides at least one in-plane bond. If both sites provide at least one such lateral bond, one of the two sites is chosen randomly and the particle hops to this site and sticks there. If none of the neighboring sites provides lateral bonds, the particle will remain at the initial adsorption site $i$.

In-plane bonds for a particle at site $i$ are present if $h_{i} + 1 \le h_{j}$ ($j = i \pm 1$). If the site provides exactly one in-plane bond ($n = 1$), it is called kink site, while a site that provides two such bonds ($n = 2$) is called a valley site. Since particles with $n \ge 1$ are immobile, the DT model represents a minimal model for MBE growth at \textit{low} $T$. In this situation, already one in-plane bond is enough to suppress particle diffusion.

In the present study, we extend the DT model by allowing adsorbed particles to explore not only nearest-neighbor lattices sites, but also sites that are further away from the deposition site. In other words, we consider the case $l \ge 1$. In general, nonequilibrium surface growth is dominated by stochastic processes that involve fluctuations not only in the deposition, but also in the diffusive motion of the particles. By setting a \textit{constant} diffusion length $l$ in the LM model, this fundamental aspect is fully neglected. Our strategy to add fluctuations to the diffusion processes in the LM model is as follows. Instead of taking a fixed diffusion length for all particles, we choose $l$ individually for each particle from a Gaussian distribution

\begin{equation}
P(l \mid l_{n}, \sigma^{2}) = \frac{1}{\sqrt{2 \pi \sigma^{2}}} \text{exp} \left[- \frac{(l-l_{n})^{2}}{2 \sigma^{2}} \right].
\end{equation}

In Eq. (8), $l_{n}$ is the mean value of the diffusion length (which we determine via KMC simulations) and the variance $\sigma^{2}$ represents the control parameter that allows to vary the degree of variability in the diffusion length $l$. In this sense, $\sigma^{2}$ controls the strength of diffusional fluctuations. The system in one dimension is illustrated in Fig. 1. We note already here that our extended model can be generalized to two dimensions. In that case, care has to be taken since there may exist multiple appropriate final sites at the same distance from the adsorption site and one has to define rules which of the possible final sites is chosen (see Sec. V).

\section{Connecting both models} 

\subsection{Strategy}

It is well established that the surface morphologies observed in MBE (and KMC) depend on the growth parameter $R(T,F)$ [see Eq. (1)]. The latter determines, in particular, the cluster properties in the submonolayer as well as the overall surface morphology in the multilayer growth regime. Our aim is to establish a direct connection between the KMC and our LM model with distributed diffusion length in order to mimic growth by the KMC model at any value of $R(T,F)$ (in the following we only use $R$). To compare the resulting morphologies in the submonolayer regime, we calculate the total number of clusters on the lattice, $N(\theta)$, and the cluster size distribution, $P(S)$, at various values of the growth parameters $R$. In the multilayer regime (see Sec. IV B), we calculate and compare layer coverages $\theta_{k}$ (with $k$ being the layer index), compute the global interface width $W(L,\theta)$ [see Eq. (13)] and we perform a scaling analysis. Moreover, we consider the height-height autocorrelation function $\Gamma(r,\theta)$ [where $r = |i - j|$, see Eq. (20)] in order to extract a correlation length $\xi_{0}$ that allows to characterize mounded surface profiles. If all these measured quantities match in both models for all values of the growth parameter $R$, we conclude that the LM model with distributed diffusion length $l$ correctly mimics the surface structures produced in KMC simulations. 

Our first main objective of this study is to find a consistent relation between the growth parameter $R$ in the KMC model, on the one hand, and the diffusion length $l$ in the LM model, on the other hand, such that the resulting morphologies are indistinguishable. Secondly, we investigate the general effect of the variance $\sigma^{2}$ in our LM model on the morphological evolution of the surface in the sub- and multilayer growth regime

\subsection{\label{relation_R_l} Diffusion properties}

\subsubsection{Nucleation length and the geometrical cluster distance}

We calculate via KMC simulations the mean-squared displacement (MSD) of adsorbed particles as function of time $\tilde{t}$ they spend on the lattice. The MSD is defined as

\begin{equation}
MSD(\tilde{t}) = \braket{\left(i(\tilde{t}) - i(0) \right)^{2}}.
\end{equation}

Here, $i(\tilde{t}) \in [1,L]$ represents the discrete position of the particle on the lattice at time $\tilde{t}$, and $i(0)$ is the site where the particle has been initially adsorbed at $\tilde{t} = 0$. Further, $\langle ... \rangle$ is an average over many realizations. Depending on the growth conditions, $MSD(\tilde{t})$ saturates at a characteristic time $\tilde{t}_{\text{S}}$ and corresponding value $MSD_{S} = MSD(\tilde{t}_{S})$. This reflects the immobilization induced by the formation of in-plane bonds. In each simulation run, only the first deposited particle is tracked since this particle is expected to travel the maximum possible distance at the given value of the growth parameter $R$. We average $MSD(\tilde{t})$ over $\cal O$ $(10^{5})$ realizations for all considered values of $R$ (see the Appendix for further details). 

From the saturation value $MSD_{S}$, we then define the ''nucleation length''

\begin{equation}
l_{n}(R) = \sqrt{MSD_{S}(R)}.
\end{equation}

An additional (and experimentally accessible) measure for the length a particle travels until getting immobilized, is the ''geometrical cluster distance''. This quantity (for a $d$-dimensional system) is given by

\begin{equation}
d_{g}(R) = \left( \frac{L^{d}}{N_{max}(R)} \right)^{1/d},
\end{equation}

where $N_{max}(R)$ is the maximum number of clusters in the first layer during submonolayer growth and $L$ is the linear system size. It is known that for one-dimensional systems with irreversible attachment, $N_{max} \sim R^{-1/4}$ \cite{Bartelt1992,Amar2004}, whereas $N_{max} \sim R^{-1/3}$ for irreversible attachment in two dimensions \cite{Amar_Family_95}.

Now the question arises whether $l_{n}$ (or $d_{g}$) might serve as an appropriate choice for the diffusion length $l$ in our LM model. To explore this issue, we plot $l_{n}$ along with $d_{g}$ as function of $R$ (in the experimentally relevant regime) in Fig. 2. For values $R < 10^{3}$, particle adsorption dominates, and we observe an increase of $l_{n}$ with $R$, while $d_{g}$ remains nearly constant. As soon as we enter the regime $R \ge 10^{3}$, particle diffusion becomes the dominant process and we identify the characteristic scaling $l_{n} \approx d_{g} \sim R^{1/4}$ ($d = 1$), because $d_{g} \sim 1/N_{max}$ [see Eq. (11)]. For the two-dimensional scenario ($d = 2$) we find $d_{g} \sim R^{1/6}$, since $d_{g} \sim (1/N_{max})^{1/2}$.

The intriguing result is that $l_{n}$ follows the \textit{same} scaling and takes (approximately) the same values. This means that it is sufficient to know $N_{max}$ (which can be experimentally determined from AFM or STM snapshots) to find both lengths, $l_{n}$ and $d_{g}$.

Based on these findings, we henceforth take the nucleation length $l_{n}$ (or, equivalently, $d_{g}$ for $R \ge 10^{3}$) as an estimate for the diffusion length $l$ in the LM model. In the following, we analyze corresponding numerical results in detail where we particularly focus on the effect of diffusional fluctuations which are controlled via $\sigma^{2}$.

\begin{figure}
	\includegraphics[width=0.95\linewidth]{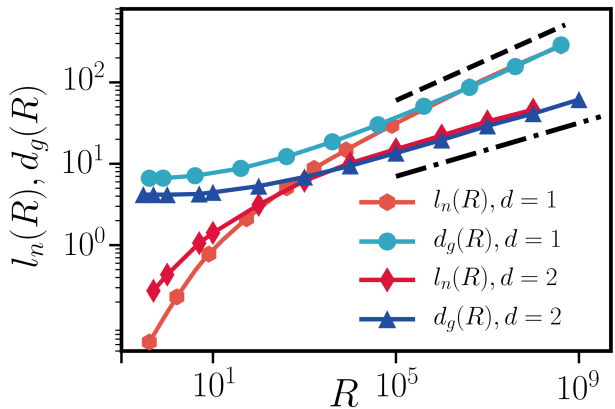}
	\caption{\label{Average cluster shape} KMC results for the nucleation length $l_{n}$ [see Eq. (10)] of particles adsorbed at the very early stage of submonolayer growth, and the geometrical distance $d_{g}$ [see Eq. (11)] between clusters, as function of $R$. The dashed black line describes the dependency $\sim R^{\gamma}$, with $\gamma = 1/4$ ($d=1$), while the dash-dotted line follows $\sim R^{1/6}$ ($d=2$).}
\end{figure}

\section{Results in one dimension}

\subsection{\label{submonolayerresults} The submonolayer growth regime}

In this section, we aim at investigating whether the LM model with mean diffusion length $l = l_{n}(R)$ and variance $\sigma^{2}$ can indeed reproduce morphologies in the submonolayer regime ($\theta < 1$) that are equivalent to those obtained from KMC simulations at arbitrary values of $R$. Here we focus on the one-dimensional case. To compare the two models quantitatively, we calculate $N(\theta)$, the number of clusters in the first layer, as well as the corresponding distribution $P(S)$ of clusters of size $S$. Since detachment of particles from cluster boundary sites is neglected, already dimers represent stable clusters. We thus distinguish between clusters $N$ (of size $2 \le S \le L$) and monomers $n$.

\subsubsection{Number of clusters in the first layer}

The evolution of $N(\theta)$ for various values of the growth parameter $R$ is shown in Fig. 3. We here focus on clusters in the first layer and monitor them up to a final coverage of $\theta = 2.5$. In this regime, we observe good agreement between the KMC (solid lines) and the LM model with constant $l$ (i.e., $\sigma^{2} = 0$ represented by dashed lines) at the lowest value of $R$ considered, $R = 4 \times 10^{2}$ (corresponding to $l = 5$ in the LM model). In particular, the location and value of the maximum, $N_{max}$, is matching perfectly. However, for larger values of $R$, we find pronounced deviations. Particularly striking are the discrepancies in $N_{max}$ and the emergence of a plateau in $N(\theta)$ within the LM model in comparison to KMC simulations with $R \ge 10^{3}$. This shows that at growth conditions where diffusion dominates over adsorption, the LM model with a constant diffusion length for all particles fails to correctly reproduce the KMC simulations. To quantify the mismatch between the two models (in absence of diffusional fluctuations, i.e., $\sigma^{2} = 0$, in the LM model), we show in Table I the values of $N_{max}$ and the relative error $\epsilon$ for various growth conditions expressed via $R$ and corresponding values of $l$.

\begin{table}
	\caption{\label{tab:example} Maxmimum number of clusters $N_{max}$ and the relative error $\epsilon$ (in $\%$) in $N_{max}$ during submonolayer growth in the LM model without fluctuations in $l$ comparison to KMC simulations at various values of the growth parameter $R$ (in $d = 1$).}
	\begin{ruledtabular}
		\begin{tabular}{lllll}
			$R$ & $l$ & $N_{max}$ KMC & $N_{max}$ LM, $\sigma^{2} = 0$ & $\epsilon$ in $\%$ \\	
			$1.5 \times 10^{2}$ & 3 & 403.85 & 403.21 & 0.16 \\	
			$4.0 \times 10^{2}$ & 5 & 333.72 & 332.56 & 0.35 \\
			$1.3 \times 10^{3}$ & 8 & 272.54 & 252.55 & 7.33 \\
			$4.0 \times 10^{3}$ & 12 & 220.08 & 186.29 & 15.35 \\
			$4.0 \times 10^{4}$ & 23 & 134.84 & 105.36 & 21.86 \\
			$8.0 \times 10^{4}$ & 30 & 116.76 & 86.99 & 25.50 \\			
			$4.0 \times 10^{5}$ & 46 & 80.69 & 59.61 & 26.13 \\
			$4.0 \times 10^{6}$ & 85 & 32.55 & 21.61 & 33.61 \\
		\end{tabular}
	\end{ruledtabular}
\end{table}

\begin{figure}
	\includegraphics[width=0.9\linewidth]{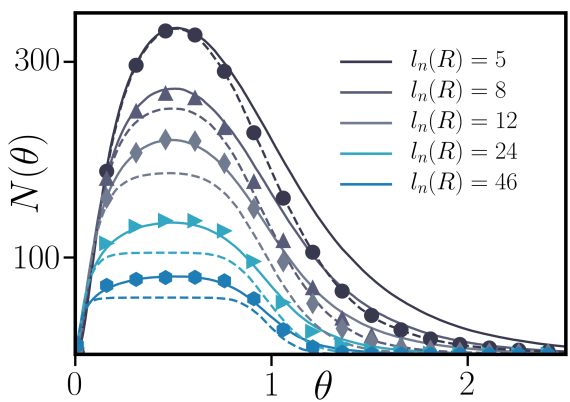}
	\caption{\label{Average cluster shape} Comparison of $N(\theta)$ in the KMC model for various values of $R$ (solid lines) with the LM model where $l = l_{n}(R)$ is used. The dashed lines represent results of the LM model with $\sigma^{2} = 0$, while the dots correspond to $N(\theta)$ with optimal values of $\sigma^{2}$.}
\end{figure}

Furthermore, Fig. 3 also shows that the deviations regarding $N(\theta)$ between the two models become much smaller when fluctuations in $l$ are ''switched on''. This is done by choosing the variance $\sigma^{2}$ appropriately for a given value of $l$ (symbols in Fig. 3 represent the LM model with optimal values of $\sigma^{2}$). To this end, we consider the difference between the maximum number of clusters
	
\begin{equation}
\Delta N_{max} = N_{max}^{KMC}(R) - N_{max}^{LM}(l,\sigma^{2})
\end{equation}
	
in the KMC and the LM model, respectively. Results for $\Delta N_{max}$ as function of $\sigma^{2}$ for various growth conditions are shown in Fig. 4. One observes that $\Delta N_{max}$ is positive for small $\sigma^{2}$, reflecting the fact that the LM model with negligible fluctuations in $l$ underestimates the values of $N_{max}$ from KMC simulations. As the strength of fluctuations is increased, $\Delta N_{max}$ decreases until it crosses the black dashed line that corresponds to $\Delta N_{max} = 0$. The values of $\sigma^{2}$ for which $\Delta N_{max} = 0$ are referred to as optimal $\sigma^{2}$ (the inset of Fig. 4 shows the dependency of the optimal value of $\sigma^{2}$ as function of $l$). These optimal values are used in Fig. 3 to match the KMC results (and they are also used for all further analyzed quantities). Upon increasing $\sigma^{2}$ above the optimal value, $\Delta N_{max}$ takes negative values. In this range, the number of clusters overshoots the values $N_{max}$ obtained from KMC simulations. Taken together, Fig. 4 shows the importance of diffusional fluctuations but also tells that their strength has to be chosen carefully.

\begin{figure}
	\includegraphics[width=0.9\linewidth]{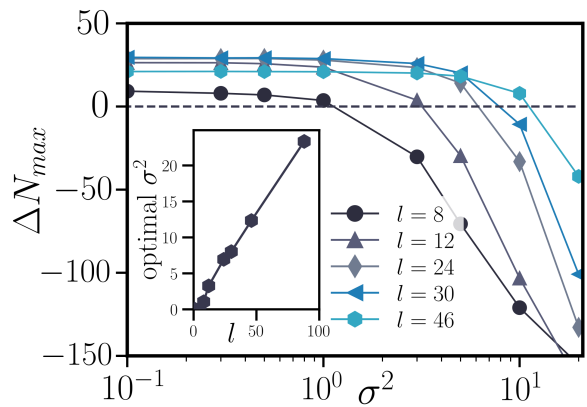}
	\caption{\label{Average cluster shape} The quantity $\Delta N_{max}$ [see Eq. (12)] as function of the variance $\sigma^{2}$ in the LM model for various values of the mean diffusion length $l_{n}(R) = l$. The black dashed line indicates $\Delta N_{max} = 0$. The inset shows the optimal values of $\sigma^{2}$ as function of $l$ fulfilling $\Delta N_{max} = 0$.}
\end{figure}

It has been shown that the asymptotic scaling of the number of clusters as function of $\theta$ (in one dimension) follows $N \sim \theta^{1/4}$ \cite{Amar2001,Tokar2008} and the results in Fig. 5 reveal that the LM model obeys this scaling only with optimal $\sigma^{2}$. Additionally, we observe unexpectedly good agreement with KMC results concerning the number of monomers $n$ as function of $\theta$ (during submonolayer growth), provided that the optimal $\sigma^{2}$ is chosen. Even though the values for $n$ obtained from KMC and LM do not perfectly match, the scaling $n \sim \theta^{-r}$, with $r \approx 0.64$ is quite similar \cite{Tokar2008}. In this context we note that the mean-field theory predicts $r = 0.5$ \cite{Amar2001} and the difference in $r$ (between simulations and mean-field theory) is because our value for $R$ is chosen too small. However, $r \approx 0.64$ was also found in \cite{Tokar2008}. This is rather surprising since we do not explicitly model the particle diffusion and thus did not expect such a resemblance. In contrast, the LM model with fixed diffusion length ($\sigma^{2} = 0$) gives a wrong scaling for both, $N$ and $n$.

\begin{figure}
	\includegraphics[width=0.9\linewidth]{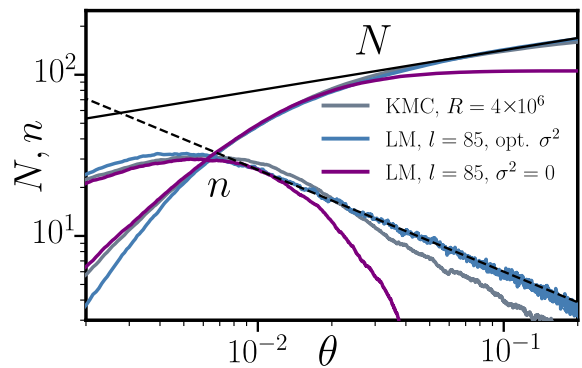}
	\caption{\label{Average cluster shape} Number of clusters $N$ and monomers $n$ as function of coverage $\theta$ on a lattice of size $L = 16384$ for the KMC model and the LM model with corresponding diffusion length and different values of the variance $\sigma^{2}$. The solid black line follows $\sim \theta^{1/4}$, while the dashed black line scales according to $\sim \theta^{-r}$, with $r \approx 0.64$ \cite{Tokar2008,Amar2001}.}
\end{figure}

\subsubsection{Cluster size distributions in the first layer}

We now consider the cluster size distribution $P(S)$ in the LM model for two representative values of the mean diffusion length, $l = 8$ and $l = 20$ (see Fig. 6). Our goal is to explore the effect of fluctuations in $l$ on $P(S)$ in the submonolayer regime (at $\theta = 0.5$) on a qualitative level only, without performing a detailed scaling analysis of cluster size distributions as done in previous studies \cite{Bartelt1992,Maass2012}. For both values of $l$, we observe a shift of the maximum of $P(S)$ towards smaller values of $S$ as the value of $\sigma^{2}$ is increased. Together with this shift, there emerges a left shoulder that indeed corresponds to the correct form of $P(S)$ for small cluster sizes $S$ in the pre-coalescence regime \cite{Michely2003,Evans2006}. At $\sigma^{2} = 0$ this shoulder is absent for $l = 24$, and too small for $l = 12$. Using the earlier obtained optimal values for $\sigma^{2}$ (see Fig. 4), we find good agreement between $P(S)$ obtained from the LM and the KMC model, respectively. As $\sigma^{2}$ is increased above the optimal $\sigma^{2}$, the maximum of $P(S)$ is shifted to smaller values of $S$ until the left shoulder vanishes and $P(S)$ becomes a monotonically decreasing function of $S$. The dependency of $P(S)$ on $\sigma^{2}$ shows that diffusional fluctuations are essential to retain the correct form of $P(S)$ in systems that model nonequilibrium surface growth with limited mobility of particles.

\begin{figure}
	\includegraphics[width=0.99\linewidth]{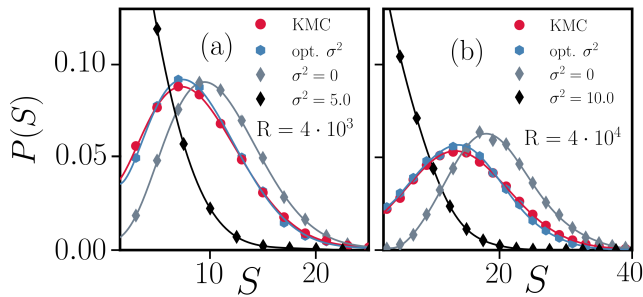}
	\caption{\label{Average cluster shape} Comparison of the cluster size distribution $P(S)$ during submonolayer growth at $\theta = 0.5$ for (a) $l = 12$, (b) $l = 24$ and various values of the variance $\sigma^{2}$ that controls the strength of diffusional fluctuations.}
\end{figure}

Finally, we shortly summarize our reasoning for the observed discrepancies in $N(\theta)$ and $P(S)$ between the KMC and the LM model without fluctuations in $l$ for certain growth conditions, $R > 4 \times 10^{2}$ ($l > 5$) and thereby justify our LM model. Nonequilibrium surface growth is dominated by stochastic processes that involve fluctuations not only in the deposition, but also in the diffusive motion of the adsorbed particles. By setting a \textit{constant} diffusion length $l$ in the LM model, this fundamental aspect is fully neglected. As a consequence, we obtain less clusters which are, moreover, too regular in size as opposed to the clusters in the KMC simulations, where the stochastic nature of diffusion is included (see Fig. 3, 5, 6). While diffusional fluctuations seem to be negligible at $R \le 4 \times 10^{2}$ ($l \le 5$), they do become significant for growth conditions where diffusion dominates, $R > 10^{3}$ ($l > 5$). Therefore, to realistically model nonequilibrium surface growth in the LM model at large values of $R$, one has to introduce diffusional fluctuations. The way we have chosen to include them is to pick the diffusion length for each deposited particle from a Gaussian distribution $P(l|l_{n}, \sigma^{2})$ where the variance controls the strength of fluctuations around the mean value $l$. 

\subsection{\label{multilayerresults} The multilayer growth regime}

Having found a suitable LM model to describe the submonolayer growth, it is an important question whether this model is also capable of describing multilayer growth. The main quantity of interest concerning the surface morphology in the multilayer regime is the global interface width \cite{Family1990,Richards1991,Chame2004,Richards1991,Barabasi1995}, defined as the root of the integrated mean square fluctuations of the local surface height at coverage $\theta$. In continuous form, the global interface width in one dimension reads

\begin{equation}
W(L,\theta) = \sqrt{\frac{1}{L} \int_{0}^{L} (h(x,\theta) - \braket{h(\theta)})^{2} \, dx}. 
\end{equation}

Here, $h(x,\theta)$ is the local surface height at position $x$ (or $i$ in discrete form) and coverage $\theta$, $L$ is the size of the substrate, and

\begin{equation}
\braket{h(\theta)} = \frac{1}{L} \int_{0}^{L} h(x,\theta) \, dx 
\end{equation}

represents the average surface height of the growing film. Thus, $W(L,\theta)$ is a measure of the surface roughness. Further, studying $W(L,\theta)$ allows to explore whether the dynamics of the growing surface exhibits universal behavior and can thus be assigned to one of the established universality classes in nonequilibrium surface growth \cite{Krug1988,Krug1993,Evans2006,Barabasi1995,Kardar1986,Bertini1997,Takeuchi2011,DasSarma1991,Lai1991,DasSarma1992}. To be more specific, investigating $W(L,\theta)$ helps to identify whether the local surface height evolves (in the hydrodynamic limit) in the functional form $\partial_{\theta} h(x,\theta) = {\cal{F}}[\nabla^{n} h(x,\theta)]$, where $\cal{F}$ is a characteristic functional involving gradient terms. Thus, examining $W(L,\theta)$ can contribute to a deeper understanding of the interface dynamics during MBE growth and may lead to improved control strategies for epitaxially fabricated devices. 

Generally, the global interface width is expected to follow the Family-Vicsek scaling relation \cite{Family_Vicsek_1985}

\begin{equation}
W(L,\theta) \sim \theta^{\beta} f\left(\frac{L}{\theta^{1/z}}\right), 
\end{equation}

where $\beta$ and $z$ are the growth and dynamic exponent, respectively. Further, $f(u)$ is a scaling function that obeys 

\begin{equation}
f(u) \sim \begin{cases} 
u^{\alpha} & u \ll 1 \\
\text{const.}& u \gg 1, 
\end{cases}
\end{equation}

which involves the global roughness exponent $\alpha = \beta z$ that depends on the two independent exponents $\beta$ and $z$. The set of these three critical exponents ($\alpha, \beta, z$) determines the universality class of the growth process under study. 

The growth exponent $\beta$ can be extracted from the short-time behavior of the interface width which is known \cite{Family_Vicsek_1985} to scale as $W(L,\theta) \sim \theta^{\beta}$ for coverages $\theta < \theta^{*}$ [with $\theta^{*}$ being the crossover coverage at which $W(L,\theta)$ reaches a saturation value $W_{sat}(L)$]. To obtain the exponents $\alpha$ and $z$, it is necessary to reach the asymptotic regime, $\theta \ge \theta^{*}$. Since the crossover coverage $\theta^{*}$ scales with system size $L$ according to \cite{Family_Vicsek_1985}

\begin{equation}
\theta^{*} \sim L^{z}, 
\end{equation}

it is very difficult to determine $\alpha$ and $z$ for large $L$. This is due to the high computational demand to reach $W_{\text{sat}}(L)$, especially when $\alpha > 1$ and $z > 2$ \cite{Punyindu1998,Punyindu2001,Punyindu2002}.

\subsubsection{Evolution of layer coverages}

In order to compare both models the initial stage of multilayer growth, we compute the evolution of the coverage in the first ten layers. In the following, layer coverages are denoted by $\theta_{k}$, with $k$ being the layer index. They are defined as 

\begin{equation}
\theta_{k} = \frac{1}{L} \sum_{i = 1}^{L} \Theta(|h_{i} - k|),
\end{equation}

with the Heaviside step function $\Theta(X)$ that obeys $\Theta(X) = 0$ for $X < 0$ and $\Theta(X) = 1$ for $X \ge 0$. We note that $\theta_{k}$ is different from the quantity $\theta$, since the latter describes the \textit{total} coverage.

Results for $\theta_{k}$ (for $k = 1$ to $k = 10$) are shown in Fig. 7 for two different values of $R$ and corresponding distributions $P(l \mid l_{n}, \sigma^{2})$. For both considered growth conditions we find perfect agreement between the layer coverage evolution in both models. To show that this agreement holds at any value of the growth parameter $R$, we present in Fig. 7 (c) the evolution of $\theta_{10}(\theta)$ for various values of $R$ and corresponding $P(l \mid l_{n}, \sigma^{2})$. Again, we find nearly perfect agreement between results from both models. Thus, we conclude that the LM model with optimal $\sigma^{2}$ yields a very good description of the KMC results during the early stages of multilayer growth.

\begin{figure}
	\includegraphics[width=1.0\linewidth]{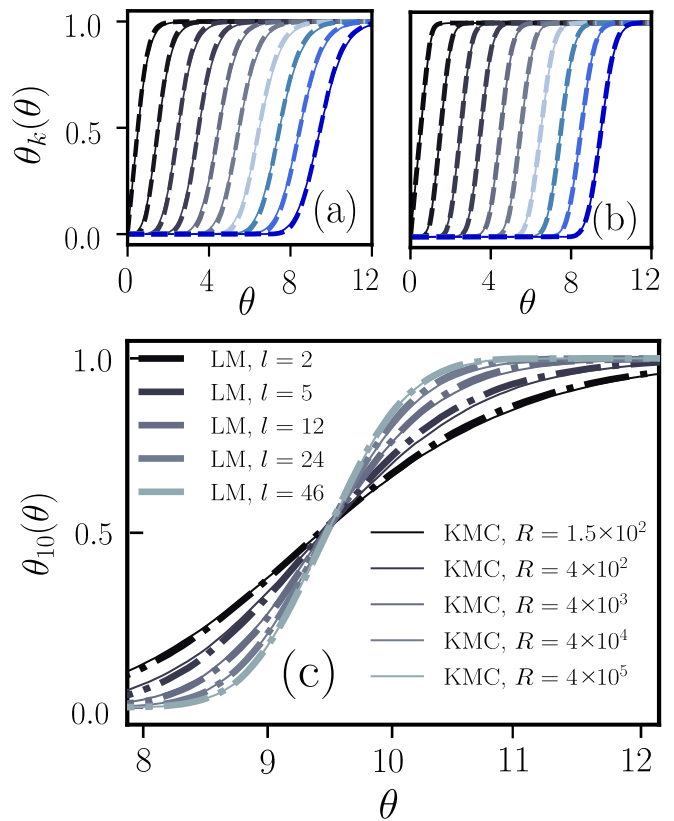}
	\caption{\label{Average cluster shape} (a) Evolution of the coverage in the first ten layers in the KMC model at $R = 10^{2}$ and $l = 5$ ($\sigma^{2} = 0.1$) in the LM model. (b) The same at $R = 4 \times 10^{4}$ in the KMC model and $l = 24$ ($\sigma^{2} = 7.0$) in the LM model. (c) Coverage evolution of the tenth layer at various growth conditions in both models. Solid lines represent KMC simulations, dotted lines are results from the LM model (with optimal variance $\sigma^{2}$).}
\end{figure}

\subsubsection{Roughness and scaling in the multilayer growth regime}

In this section we study the regime of many layers (up to $\theta = 10^{6}$) by investigating the global interface width $W(L,\theta)$ [see Eq. (8)]. The evolution of $W(L,\theta)$ for different system sizes $L$ and four exemplary values of $R$ and corresponding distributions $P(l \mid l_{n}, \sigma^{2})$ is shown in Fig. 8. Results from KMC simulations are given by symbols, while for the LM model, $W(L,\theta)$ is represented by solid lines.

\begin{figure}
	\includegraphics[width=1.0\linewidth]{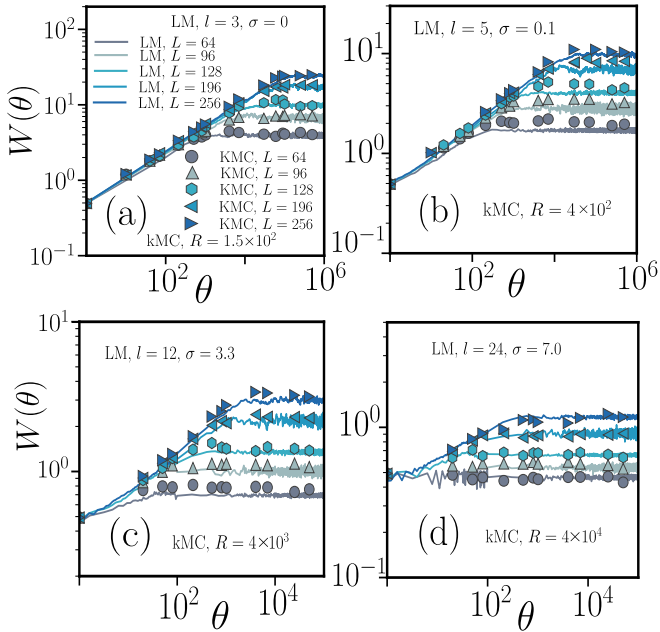}
	\caption{\label{Average cluster shape} Global interface width $W(L,\theta)$ as function of coverage $\theta$ for four different values of $R$ in the KMC model (symbols) and corresponding diffusion length $l$ in the LM model (solid lines) with optimal values for $\sigma^{2}$ for system sizes from $L = 64$ to $L = 256$. }
\end{figure}

According to the Family-Vicsek scaling relation \cite{Family_Vicsek_1985}, $W(L,\theta)$ initially shows power-law scaling, $W(L,\theta) \sim \theta^{\beta}$. From our KMC data, we identify $\beta \approx 1/3$ for all considered values of $R$. The reason why $\beta$ does not depend on $R$ is that varying $R$ does not change the symmetry properties of the system \cite{Krug1993,Burioni1995,Park2001}. The growth exponent $\beta \approx 1/3$ also correctly describes the roughening in the LM model. Not only the scaling exponent $\beta \approx 1/3$ is the same in both models, but also the actual values of $W(L,\theta)$ for all considered growth conditions and system sizes $L$ [see Fig. 9 (c)].

\begin{figure}
	\includegraphics[width=1.0\linewidth]{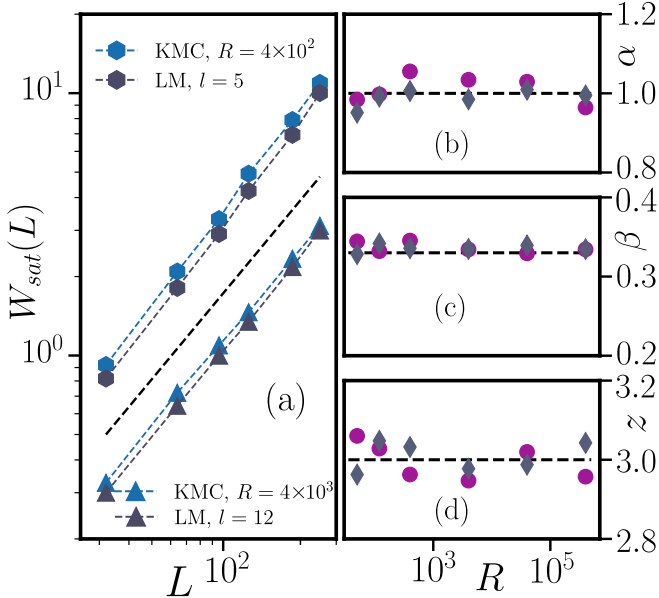}
	\caption{\label{Average cluster shape} (a) Scaling of the saturation roughness $W_{sat}$ as function of system size $L$ for two values of $R$ in the KMC model and the corresponding values of $l$ and $\sigma^{2}$ in the LM model (dotted black line $\sim L^{1}$). (b) Roughness exponent $\alpha$, (c) growth exponent $\beta$ and (d) dynamic exponent $z$ for various growth conditions in both models. The lines in (b)-(d) represent the values of $\alpha, \beta, z$ according to the VLDS universality class in one dimension.}
\end{figure}

After a crossover to the asymptotic regime at the cross-over coverage $\theta^{*}$, the interface width saturates. Independent of the value of $R$ and $P(l \mid l_{n}, \sigma^{2})$, the onset of roughness saturation scales $\theta^{*} \sim L^{3}$. Again, the scaling exponents are very similar in both models for all considered growth conditions [See Fig. 9 (d)].

In addition, we find that the saturation values obey $W_{sat}(L) \sim L^{1}$ [See Fig. 9 (b)]. Thus, the roughness exponent $\alpha \approx 1$ is the same in both models for all considered growth conditions.

At this point it is worth to recall that our LM model involves a fluctuating diffusion length. To demonstrate the importance of fluctuations in the multilayer regime, we show in Fig. 10 the relative error $\epsilon(W_{sat}) = |W_{sat}(L)^{LM} - W_{sat}(L)^{KMC}| / W_{sat}(L)^{KMC} \times 100$ of $W_{sat}(L)$ for the LM model with $\sigma^{2} = 0$ and the version with optimal $\sigma^{2}$. We find that, as $L$ is increased, $\epsilon(W_{sat})$ diminishes with optimal $\sigma^{2}$, while the error is $\ge 40 \%$ in absence of diffusional fluctuations ($\sigma^{2} = 0$).

\begin{figure}
	\includegraphics[width=1.0\linewidth]{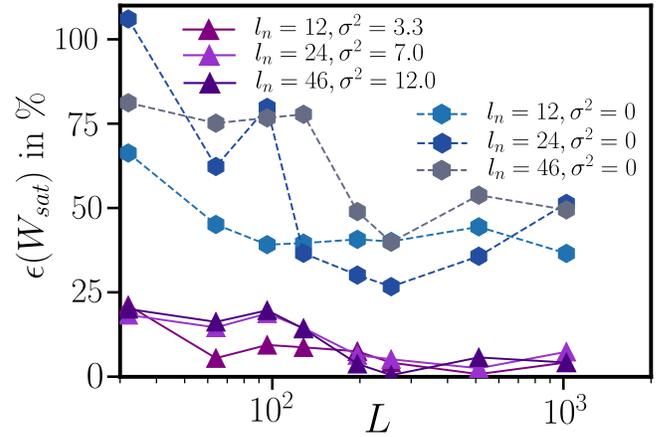}
	\caption{\label{Average cluster shape} Relative error $\epsilon (W_{sat})$ of the saturation roughness $W_{sat}(L)$ in the LM model compared to KMC simulations. Results for $\epsilon (W_{sat})$ without fluctuations in the diffusion length $l$ (i.e., $\sigma^{2} = 0$) are given by hexagons, while $\epsilon (W_{sat})$ with optimal $\sigma^{2}$ are represented by triangles.}
\end{figure}

We now turn back to the scaling behavior. In summary, we identify the following critical exponents in our simulations: $\alpha \approx 1$, $\beta \approx 1/3$, $z \approx 3$. As known from simulations and analytical calculations of the DT model (with $l=1$), this set of critical exponents belongs to the Villain-Lai-Das Sarma (VLDS) universality class in one dimension \cite{Lai1991, DasSarma1992,Reis2004}. The corresponding evolution equation for the surface height in the hydrodynamic limit is given by

\begin{equation}
\partial_{t} h(x,t) = -\nu_{4} \partial_{x}^{4} h(x,t) + \lambda_{4} (\partial_{x}^{2} h(x,t))^{2} + F.
\end{equation}

In Eq. (19), $h(x,t)$ is the surface height at position $x$ at time $t$, $\nu_{4}$ and $\lambda_{4}$ are constants, and $F$ is a Gaussian white noise, representing the randomness accompanying the deposition of particles. Thus, Eq. (19) is a stochastic, nonlinear partial differential equation. Note that for $\lambda_{4} = 0$ the equation reduces to the linear Mullins-Herring (MH) equation (characterized by the critical exponents, in one dimension, $\alpha = 3/2$, $\beta = 3/8$, $z = 4$) \cite{Herring1950, Mullins1957}. The non-linear equation, $\lambda_{4} \neq 0$, is known to have the same symmetry as several discrete lattice models (including the DT model) \cite{DasSarma1991,DasSarma1992} that are frequently used to model surface growth. Thus, the nonlinear equation displays the same set of critical exponents.

As known from experiments \cite{Neave1983,Alexandre1985,Kunkel1990} and KMC simulations \cite{Smilauer1993,Assis2013,Assis2015}, the surface of a growing thin film becomes smoother as the value of $R$ is increased. We systematically study this smoothing of the surface and the thus resulting decrease of the surface roughness in the LM model in detail by plotting $W_{sat}$ versus $l$ for different system sizes $L$ in Fig. 11 (a). We observe that $W_{sat}(L,l)$ obeys a power-law, $W_{sat}(L,l) \sim l^{-\phi}$, with scaling exponent $\phi \approx 3/2$ that is independent of $L$. In KMC simulations, the saturation roughness decreases according to $W_{sat}(L,R) \sim R^{-\delta}$, with $\delta \approx 1/2$ [see Fig. 11 (b)]. To confirm the correctness of the scaling exponents $\phi$ and $\delta$, we define a rescaled saturation roughness for both models, $W_{sat}^{RE} = W_{sat}(L,l) / (L^{\alpha} l^{-\phi})$ (LM) and $W_{sat}^{RE} = W_{sat}(L,R) / (L^{\alpha} R^{-\delta})$ (KMC). Results for $W_{sat}^{RE}$ are shown in Fig. 11 (c). We find that $W_{sat}^{RE}$ as function of $l$ (LM model) is indeed a constant. The same holds for KMC simulations where $W_{sat}^{RE}$ is plotted as function of $l_{n}(R)$.

\begin{figure}
	\includegraphics[width=1.0\linewidth]{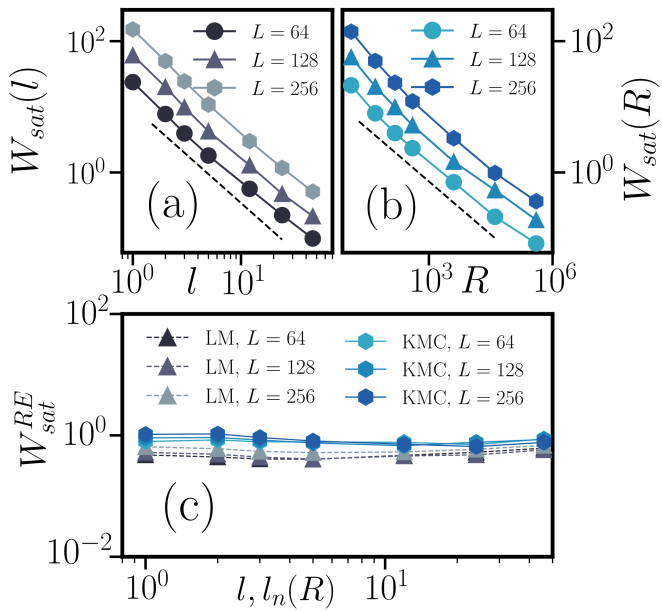}
	\caption{\label{Average cluster shape} (a) Scaling of the saturation roughness $W_{sat}(l)$ in the LM model as function of the average diffusion length $l$ with optimal values for $\sigma^{2}$ for different system sizes $L$. The black dotted line follows $\sim l^{-3/2}$. (b) Scaling of $W_{sat}(R)$ as function of the growth parameter $R$. The black dotted line follows $\sim R^{-1/2}$. (c) Rescaled saturation roughness $W_{sat}^{RE}$ in the LM model and corresponding KMC simulations.}
\end{figure}

\subsubsection{Interface profiles and the effect of diffusional fluctuations}

Growth instabilities can induce the formation of mound-like patterns and it is well accepted that the original DT model displays quasiregular mound formation \cite{Punyindu2001,DasSarma2000,DasSarma2002,Kanjanaput2010,Luis2017,Luis2019,Pereira2019}. To investigate how diffusional fluctuations alter this characteristic feature, we show exemplary interface profiles for two diffusion lengths $l$ and different values of $\sigma^{2}$ in Fig. 12 (a) - (f) [please note that $h(x) = h_{i} - h_{min}$, where $h_{min}$ is the minimum height in the depicted profiles in Fig. 12]. Upon increase of $\sigma^{2}$ from zero, the characteristic mound size decreases, while the number of mounds increases. Further, individual mounds appear to be sharper and steeper, with the result that the overall interface looks rougher, yielding a higher value of the interface width. 

To analyze in detail how the value of $\sigma^{2}$ modifies the interface height profile, we calculate a characteristic length $\xi_{0}$ that contains information regarding the characteristic mound size \cite{To2018,Luis2019}. This quantity is defined as the first zero of the height-height correlation function

\begin{equation}
\Gamma(r,\theta) = \frac{ \langle \tilde{h}_{i}(\theta) \tilde{h}_{i+r} (\theta) \rangle}{ \langle \tilde{h}_{i}(\theta)^{2} \rangle},
\end{equation}

with $\tilde{h}_{i} (\theta) = h_{i} (\theta) - \braket{h(\theta)}$. Calculating $\Gamma(r) / \Gamma(0)$ reveals that with increasing $\sigma^{2}$, $\Gamma(r) / \Gamma(0)$ decays faster to zero with the distance $r = |i - (i+r)|$ from site $i$. As a consequence, the value of $\xi_{0}$ decreases [see Fig. 12 (g) and (h) where $\Gamma(r) / \Gamma(0)$ reaches zero at smaller values of $r$ as $\sigma^{2}$ is increased]. This goes along with a decrease of the characteristic mound size.

\begin{figure}
	\includegraphics[width=1.0\linewidth]{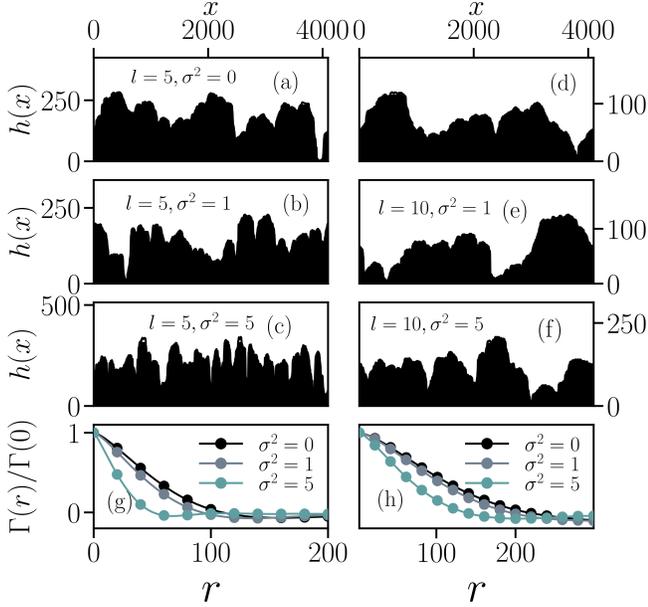}
	\caption{\label{Average cluster shape} (a) - (f) Height profiles in the LM model at $\theta = 10^{6}$ for two representative values of $l$ and different values of the variance $\sigma^{2}$ at coverage $\theta = 10^{6}$ in systems of size $L = 4096$. (g) Height-height correlation function $\Gamma(r) / \Gamma(0)$ for the depicted system settings with $l = 5$ and different values of $\sigma^{2}$ at $\theta = 10^{4}$. (h) The same as in (g) for $l = 10$.}
\end{figure}

\begin{figure}
	\includegraphics[width=0.95\linewidth]{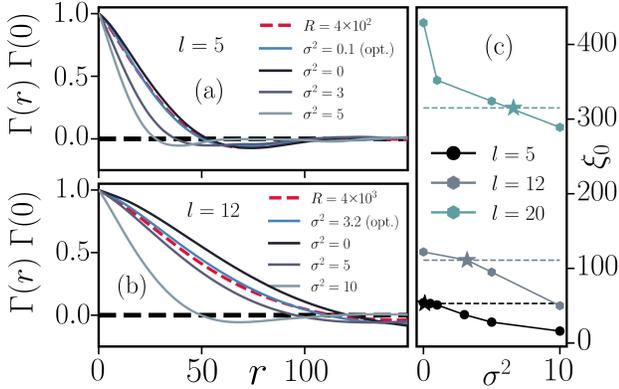}
	\caption{\label{Average cluster shape} (a) Height-height correlation function $\Gamma(r) / \Gamma(0)$ for KMC simulations at $R = 4 \times 10^{2}$ together with results of the LM model at $l = 5$ and various values of the variance $\sigma^{2}$. (b) The same as in (a) for $R = 4 \times 10^{3}$ and $l = 12$. (c) Correlation length $\xi_{0}$ for different combinations of $l$ and $\sigma^{2}$ (lines with symbols) together with corresponding values of $\xi_{0}$ from KMC simulations (dashed lines). Stars in (c) represent the optimal values of $\sigma^{2}$.}
\end{figure}

To demonstrate the equality of the morphologies generated by both models, we compare $\Gamma(r) / \Gamma(0)$ obtained from the KMC and the LM model at $\theta = 10^{4}$ in Fig. 13. For both considered growth conditions [$l = 5$ and $l = 12$, Fig. 13 (c) additionally shows $\xi_{0}$ for $l = 20$], we observe good agreement using the optimal values of $\sigma^{2}$.

\section{Results in two-dimensions}

The analysis so far was restricted to systems in one dimension. From a physical point of view, however, it is clear that the case of two spatial dimensions is more relevant. The aim of this section is to show by exemplary calculations that the mapping strategies developed in the one-dimensional case also work in two dimensions. To start with, we note that the procedure to relate the values of $R$ in KMC simulations to the parameters $l$ and $\sigma^{2}$ in the LM model in two dimensions can be followed as in the one-dimensional case. However, the decision for the final attachment site has to be carefully considered, since in two dimensions more than two lattice sites at the same distance from the adsorption site can provide at least one lateral bond. As a first step, we decided to select the cluster boundary site for attachment that is closest to the initial adsorption site and in case their exist multiple at the same distance, we randomly choose one of them. However, we want to emphasize that different variants for choosing the final site are possible and even necessary when turning towards growth conditions where edge-diffusion and bond breaking is possible.

\subsection{The submonolayer growth regime}

To demonstrate that our approach also works on two-dimensional lattices, we show in Fig. 14 the maximum number of clusters $N_{max}$ in the submonolayer growth regime as function of $R$ and corresponding $P(l | l_{n}, \sigma^{2})$ (with optimal $\sigma^{2}$). The results for $N_{max}$ reflect good agreement between both models for all considered growth conditions. In particular, $N_{max}$ decays identically in both models as the values of $R$ and $l$ are increased. 

Analyzing $P(S)$ we observe, analogous to the one-dimensional scenario (see Fig. 6 and Sec. IV. A. 2.), a shift of the peak of $P(S)$ towards smaller values of $S$ as the strength of diffusional fluctuations is increased (see Fig. 15). This results in the emergence of a left shoulder in $P(S)$ as observed earlier in one dimension. For too large values of $\sigma^{2}$ the distributions $P(S)$ become monotonically decreasing functions of $S$, which is clearly unphysical. For $l = 10$, this behavior is seen already at $\sigma^{2} = 5$, while for $l = 30$ it only occurs for $\sigma^{2} > 10$.

\begin{figure}
	\includegraphics[width=0.95\linewidth]{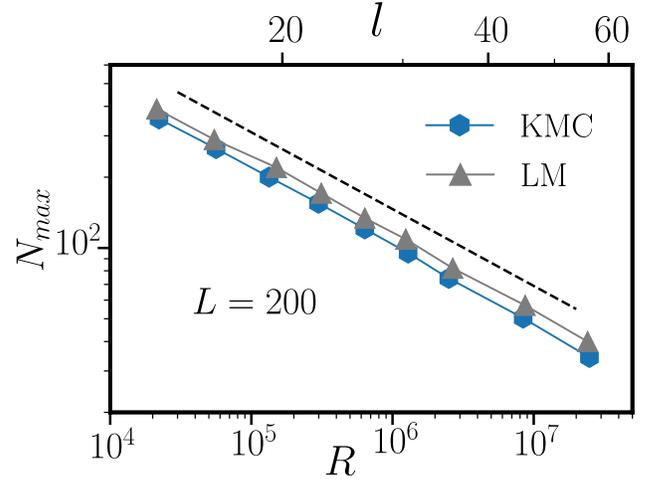}
	\caption{\label{Average cluster shape} $N_{max}$ in the submonolayer growth regime on two dimensional substrates of lateral size $L = 200$ for various growth conditions in the KMC and the LM model, respectively. The black dotted line follows $\sim R^{-1/3}$.}
\end{figure}

\begin{figure}
	\includegraphics[width=0.95\linewidth]{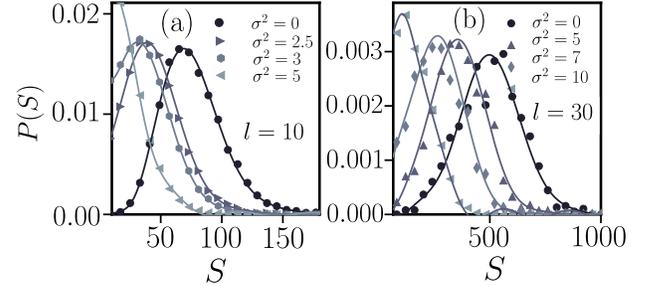}
	\caption{\label{Average cluster shape} Cluster size distribution $P(S)$ during submonolayer growth at $\theta = 0.3$ on two dimensional substrates of lateral size $L = 200$ for two values of $l$ and various strengths of diffusional fluctuations controlled via $\sigma^{2}$ in the LM model.}
\end{figure}

\subsection{The multilayer growth regime}

To compare both models on two-dimensional lattices in the multilayer regime, we show exemplary lattice structures in Fig. 16 for two values of $R$ and corresponding optimal distributions $P(l|l_{n}, \sigma^{2})$. While the lattice structures at $R = 4 \times 10^{2}$ look indistinguishable [Fig. 16 (a) and (b)], we find visible deviations at $R = 2 \times 10^{4}$ [Fig. 16 (d) and (e)]. These discrepancies may be resolved by using a different variant for choosing the final attachment site as discussed earlier in this section. Despite the spatial deviations, the functions $\Gamma(r) / \Gamma(0)$ [see Eq. (20) and plots in Fig. 16 (c) for $R = 4 \times 10^{2}$ and Fig. 16 (f) for $R = 2 \times 10^{4}$] reveal a good agreement between both models concerning height-height correlations and the correlation length $\xi_{0}$. Both of these quantities are very sensitive to changes in $l$ and $\sigma^{2}$, as shown in Fig. 12 and Fig. 13 for the one-dimensional case.

\begin{figure}
	\includegraphics[width=0.95\linewidth]{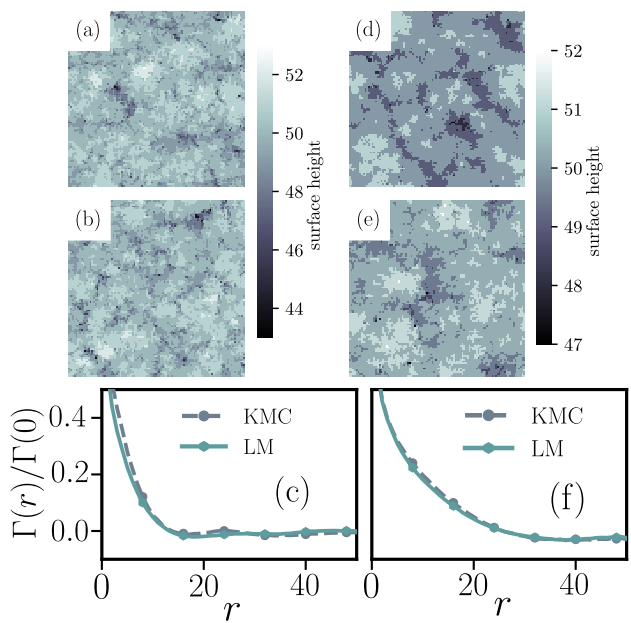}
	\caption{\label{Average cluster shape} (a) Exemplary lattice structures (of size $100 \times 100$ from lattice with $L = 200$) from KMC simulations with $R = 4 \times 10^{2}$ (a) and $R = 2 \times 10^{4}$ (d) at $\theta = 50$ along with results obtained from the LM model with $l = 4$ and $\sigma^{2} = 1.1$ (b) $l = 11$ and $\sigma^{2} = 2.8$ (e) at the same coverage, $\theta = 50$. (c) and (f) depict $\Gamma(r) / \Gamma(0)$ for both considered growth conditions in the KMC and the LM model, respectively.}
\end{figure}

\section{Conclusions and Outlook}

In this work, we have introduced an extended limited mobility (LM) model for nonequilibrium surface growth, which is capable of predicting \textit{low} temperature MBE growth for arbitrary values of the growth parameter $R$. Compared to earlier versions of the LM model, particularly the DT model, our extension concerns the diffusion length $l$ which we treat as a variable parameter whose value for each deposited particle is chosen from a Gaussian distribution.

To relate our LM model to another standard model for surface growth, namely KMC, we proposed to set the mean value of $l$ equal to the nucleation length $l_{n}$ resulting from short KMC simulations for the particle displacements. We tested this ansatz by comparing LM and KMC results for the cluster evolution during sub- and multilayer growth. While the LM model with fixed $l$ works well at small values of the growth parameter $R$, this is not the case at larger $R$. As a next step, we therefore included fluctuations to the diffusion length of particles in the LM model in order to model diffusional fluctuations. Specifically, we employed a Gaussian distribution where the mean diffusion length $l$ is given by the nucleation length extracted from KMC, whereas the variance $\sigma^{2}$ is fitted to match the maximum number of clusters $N_{max}$ in KMC simulations during growth in the submonolayer regime. For each considered value of $l$ we have identified the variance $\sigma^{2}$ that leads to $N_{max}^{KMC} - N_{max}^{LM} \approx 0$. Using these \textit{optimal} values of $\sigma^{2}$ also lead to nearly identical cluster size distributions $P(S)$.

Turning towards multilayer growth, we compared layer coverages for different growth conditions and found excellent agreement between both models. Moreover, we analyzed in detail the global interface width for different system sizes up to coverages deep in the regime of saturated surface roughness. Not only is transient regime of the global interface width identical in both models, but also the crossover coverage where saturation is reached. Additionally, we showed that by using our LM model with variable diffusion length, also the values of the saturation roughness match in both models for all considered system sizes and growth conditions. A scaling analysis revealed that the LM model belongs to the VLDS universality class for arbitrary diffusion lengths. We also observed that the variance $\sigma^{2}$ can strongly alter the interface height profile in the high coverage regime. As the value of $\sigma^{2}$ is increased, we observed less and, at the same time, steeper mounds. Moreover, we found good agreement concerning height-height correlations in both models using the optimal values of $\sigma^{2}$ in both, one- and two-dimensional systems.

The present model can be extended in various directions. First of all, it is possible to modify the model such that it also mimics MBE growth at \textit{high} temperatures where detachment of particles is present. This may be achieved by using a mixture of the transition rules of the DT and the Wolf-Villain model \cite{Wolf1990,Villain1991} with a variable, distributed diffusion length. Concerning the two-dimensional scenario, it would be very interesting to investigate how different variants for the rules regarding the final attachment site of deposited particles affects the cluster shape properties and the overall growth behavior. This is especially important when moving towards higher temperatures or lower binding energies, where clusters are usually compact rather than ramified.

Second, the effect of an additional energy barrier for interlayer diffusion processes across step-edges, usually referred to as Ehrlich-Schwoebel barrier, can be included to account for growth instabilities. Normally, in presence of such a barrier, KMC simulations are slowed down due to the sampling of diffusion trajectories of free particles on top of clusters that are reflected at the cluster edge due to the additional energy-barrier for crossing step-edges. A physically reasonable treatment of an Ehrlich-Schwoebel barrier in our LM model would lead to a further computational speedup compared to KMC simulations.

Finally, we want to point out that, especially concerning growth conditions where the critical cluster size takes large values, there exist alternative numerical techniques beyond the lattice-based models with limited particle mobility that can be further advanced to realistically model this specific growth regime. Examples include level-set \cite{Vvedensky2002} and geometry-based \cite{Evans2003} models. 

\begin{acknowledgments}
	This work was funded by the Deutsche Forschungsgemeinschaft (DFG, German Research Foundation) - Projektnummer A7 - SFB 951.
\end{acknowledgments}

\bibliography{basename of .bib file}

\begin{thebibliography}{10}

\bibitem{Yao2016}

M.~Y.~Yao, F.~Zhu, C.~Q.~Han, D.~D.~Guan, C.~Liu, D.~Qian, and J.~F.~Jia, Sci. Rep. \textbf{6}, 213226 (2016).

\bibitem{Wofford2017}

J.~M.~Wofford, S.~Nakhaie, T.~Krause, X.~Liu, M.~Ramsteiner, M.~Hanke, H.~Riechert, and J.~M.~J.~Lopes, Sci. Rep. \textbf{7}, 43644 (2017).

\bibitem{Jin2016}

L.~Jin, D.~Zhang, H.~Zhang, J.~Fang, Y.~Liao, T.~Zhou, C.~Liu, Z.~Zhong, and V.~G.~Harris, Sci. Rep. \textbf{6}, 34030 (2016).

\bibitem{Tobin1990}

P.~S.~Tobin, S.~M.~Vernon, C.~Bajgar, S.~Wojtczuk, M.~R.~Melloch, A.~Keshavarzi, T.~B.~Stellwag, S.~Venkatensan, M~Lundstrom, and K.~A.Emery, IEEE Trans. Electron Dev. \textbf{37}, 469-477 (1990).

\bibitem{Barabasi1995}

A.~L.~Barabasi and H.~E.~Stanley, Fractal Concepts in Surface Growth, Cambridge University Press (1995).

\bibitem{Pimpinelli1998}

A.~Pimpinelli and J.~Villain, Physics of Crystal Growth, Cambridge University Press (1998).

\bibitem{Michely2003}

T.~Michely and J.~Krug, Islands, Mounds and Atoms, Springer (2003).

\bibitem{Evans2006}

J.~W.~Evans, P.~A.~Thiel, and M.~C.~Bartelt, Surf. Sci. Rep. \textbf{61}, 1 (2006).

\bibitem{Vvedensky2002}

C.~Ratsch, M.~F.~Gyure, R.~E.~Calfisch, F.~Gibou, M.~Petersen, M.~Kang, J.~Garcia and D.~D.~Vvedensky, Phys. Rev. B \textbf{65}, 195403 (2002).

\bibitem{Osher2001}

S.~Chen, B.~Merriman, M.~Kang, R.~E.~Calfisch, C.~Ratsch, L.-T.~Cheng, M.~Gyure, R.~P.~Fedkiw, C.~Anderson and S.~Osher J. Comput. Phys. \textbf{167}, 475-500 (2001).

\bibitem{Evans2003}

M.~Li, M.~C.~Bartelt and J.~W.~Evans, Phys. Rev. B \textbf{68}, 121401(R) (2003).

\bibitem{Spraque1991}

C.~M.~Gilmore and J.~A.Spraque, Phys. Rev. B \textbf{44}, 8950 (1991).

\bibitem{Gilmore1992}

C.~M.~Gilmore, Journal of Vacuum Science und Technology A \textbf{10}, 1597 (1992).

\bibitem{Spraque1993}

C.~M.~Gilmore and J.~A.Spraque, Nanostructured Materials \textbf{2}, 301-310 (1993)

\bibitem{Moseler1995}

H.~Haberland, Z.~Insepov and M.~Moseler, Phys. Rev. B \textbf{51}, 11061 (1995)

\bibitem{Srolovitz1996}

L.~Dong, R.~W.~Smith, and D.~J.~Srolovitz, J. Appl. Phys. \textbf{80}, 5682 (1996)

\bibitem{Ferrando2000}

F.~Baletto, C.~Mottet, and R.~Ferrando, Surface Science \textbf{446}, 31-45 (2000)

\bibitem{Burgos2005}

E.~B.~Halac, M.~Reinoso, A.~G.~Dall'As\'en, and E.~Burgos, Phys. Rev. B \textbf{71}, 115431 (2005)

\bibitem{Chirita2016}

D.~Edstr\"om, D.~G.~Sangiovanni, L.~Hultman, I.~Petrov, J.~E.~Greene and V.~Chirita, Journal of Vacuum Science und Technology A \textbf{34}, 041509 (2016).

\bibitem{Haselwandter2005}

A.~L.-S.~Chua, C.~A.~Haselwandter C.~Baggio, and D.~D.~Vvedensky, Phys. Rev. E \textbf{72}, 051103 (2005).

\bibitem{Haselwandter2006}

C.~A.~Haselwandter and D.~D.~Vvedensky, Phys. Rev. B \textbf{74}, 121408(R) (2006).

\bibitem{Haselwandter2007_1}

C.~A.~Haselwandter and D.~D.~Vvedensky, Phys. Rev. Lett. \textbf{98}, 046102 (2007).

\bibitem{Haselwandter2007_2}

C.~A.~Haselwandter and D.~D.~Vvedensky, Europhys. Lett. \textbf{77}, 38004 (2007).

\bibitem{Haselwandter2008}

C.~A.~Haselwandter and D.~D.~Vvedensky, Phys. Rev. E \textbf{77}, 061129 (2008).

\bibitem{Haselwandter2010}

C.~A.~Haselwandter and D.~D.~Vvedensky, Phys. Rev. E \textbf{81}, 021606 (2010).

\bibitem{Weeks1979}

J.~D.~Weeks and G.~H.~Gilmer, Advances in Chemical Physics \textbf{40} (1979).

\bibitem{Maksym1988}

P.~A.~Maksym, Semicond. Sci. Technol. \textbf{3}, 594 (1988).

\bibitem{Kotrla1996}

M.~Kotrla, Computer Physics Communications \textbf{97}, 82-100 (1996).

\bibitem{Levi1997}

A.~C.~Levi and M.~Kotrla, J. Phys.: Condens. Matter \textbf{9}, 299 (1997).

\bibitem{Martynec2018}

T.~Martynec and S.~H.~L.~Klapp, Phys. Rev. E \textbf{98}, 042801 (2018).

\bibitem{Kleppmann2015_1}

N.~Kleppmann, and S.~H.~L.~Klapp, Phys. Rev. B \textbf{91}, 045436 (2015).

\bibitem{Kleppmann2015_2}

N.~Kleppmann, and S.~H.~L.~Klapp, J. Chem. Phys. {\bf 142}, 064701 (2015).

\bibitem{Kleppmann2016}

N.~Kleppmann, and S.~H.~L.~Klapp, Phys. Rev. B {\bf 94}, 241404(R) (2016).

\bibitem{Klopotek_I} 

M.~Oettel, M.~Klopotek, M.~Dixit, E.~Empting, T.~Schilling, and H.~Hansen-Goos, J. Chem. Phys. {\bf 145}, 074902 (2016).

\bibitem{Klopotek_II} 

M.~Klopotek, H.~Hansen-Goos, M.~Dixit, T.~Schilling, F.~Schreiber, and M.~Oettel, J. Chem. Phys. {\bf 146}, 084903 (2017).

\bibitem{Jana2013}

P~K.~Jana, and A.~Heuer, J. Chem. Phys. {\bf 138}, 124708 (2013).

\bibitem{ClarkeVvedensky98} 

S.~Clarke and D.~D.~Vvedensky, J. Appl. Phys. \textbf{63}, 2272 (1988).

\bibitem{Evans1995}

M.~C.~Bartelt and J.~W.~Evans, Phys. Rev. Lett. \textbf{75}, 4250 (1995).

\bibitem{Family1996}

J.~G.~Amar and F.~Family, Phys. Rev. B \textbf{54}, 14742 (1996).

\bibitem{Hohage1996}

M.~Hohage, M.~Bott, M.~Morgenstern, Z.~Zhang, T.~Michely, and G.~Cosma, Phys. Rev. Lett. \textbf{76}, 2366 (1996).

\bibitem{Evans2002}

K.~J.~Caspersen, A.~R.~Layson, C.~R.~Stoldt, V.~Fournee, P.~A.~Thiel, and J.~W.~Evans, Phys. Rev. B \textbf{65}, 193407 (2002). 

\bibitem{Evans2008}

M.~Li, P.-W.~Chung, E.~Cox, C.~J.~Jenks, P.~A.~Thiel, and J.~W.~Evans, Phys. Rev. B \textbf{77}, 033402 (2008). 

\bibitem{Kratzer2002}

P.~Kratzer and M.~Scheffler, Phys. Rev. Lett. \textbf{88}, 036102 (2002).

\bibitem{Godbey1994}

D.~J.~Godbey, J.~V.~Lill, and J.~Deppe, Appl. Phys. Lett. \textbf{65}, 711 (1994).

\bibitem{Nie2017}

Y.~Nie, C.~Liang, P.~R.~Cha, L.~Colombo, R.~M.~Wallace and K.~Cho, Sci. Rep. \textbf{7}, 2977 (2017).

\bibitem{Krause2004}

B.~Krause, F.~Schreiber, H.~Dosch, A.~Pimpinelli and O.~H.~Seeck, Europhys. Lett. \textbf{65}, 372-378 (2004)

\bibitem{Choudhary2006}

D.~Choudhary, P.~Clancy, R.~Shetty and F.~Escobedo, Adv. Funct. Mater. \textbf{16}, 1768-1775 (2006)

\bibitem{Bommel2014}

S.~Bommel, N.~Kleppmann, C.~Weber, P.~Sch\"afer, J.~Novak, S.~V.~Roth, F.~Schreiber, S.~H.~L.~Klapp, and S.~Kowarik, Nat. Comm. \textbf{5}, 5388 (2014).

\bibitem{Acevedo2016}

Y.~M.~Acevedo, R.~A.~Cantrell, P.~G.~Berard, D.~L.~Koch and P.~Clancy, Langmuir, \textbf{32}, 3045-3056 (2016)

\bibitem{Schoell2003}

L.~Mandreoli, J.~Neugebauer, R.~Kunert and E.~Sch\"{o}ll, Phys. Rev. B \textbf{68}, 155429 (2003).

\bibitem{Smereka2005}

J.~P.~DeVita, L.~M.~Sander and P.~Smereka, Phys. Rev. B \textbf{72}, 205421 (2005).

\bibitem{Opplestrup2006}

T.~Opplestrup, V.~V.~Bulatov, G.~H.~Gilmer, M.~H.~Kalos, and B.~Sadigh, Phys. Rev. Lett. \textbf{97}, 230602 (2006).

\bibitem{Chou2006}

C.-C.~Chou and M.~L.~Falk, J. Comput. Phys. \textbf{217}, 519 (2006)

\bibitem{Tokar2008}

V.~I.~Tokar and H.~Dreyss\'{e}, Phys. Rev. E \textbf{77}, 066705 (2008).

\bibitem{Tokar2009}

V.~I.~Tokar and H.~Dreyss\'{e}, Phys. Rev. B \textbf{80}, 161403 (R) (2009).

\bibitem{Tsalikis2013}

D.~G.~Tsalikis, C.~Baig, V.~G.~Mavrantzas, E.~Amanatides, and D.~Mataras, J. Chem. Phys. \textbf{139}, 204706 (2013)

\bibitem{Family1990}

F.~Family, Physica A, \textbf{168}, 561-580 (1990).

\bibitem{Wolf1990}

D.~E.~Wolf and J.~Villain, Europhys. Lett. \textbf{13}, 389 (1990).

\bibitem{Villain1991}

J.~Villain, J. Phys. I \textbf{1}, 19 (1991).

\bibitem{DasSarma1991}

S.~Das Sarma and P.~Tamborenea, Phys. Rev. Lett. \textbf{66}, 325 (1991).

\bibitem{DasSarma1992}

S.~Das Sarma and S.~V.~Ghaisas, Phys. Rev. Lett. \textbf{69}, 3762 (1992).

\bibitem{DasSarma2002}

S.~Das Sarma, P.~Punyindu Chatraphorn, and Z.~Toroczkai, Phys. Rev. E \textbf{65}, 036144 (2002).

\bibitem{DasSarma2000}

S.~Das Sarma, P.~Punyindu Chatraphorn, and Z.~Toroczkai, Surf. Sci. Lett. \textbf{457}, L369-L375 (2000).

\bibitem{Kanjanaput2010}

W.~Kanjanaput, S.~Limkumnerd and P.~Chatraphorn, Phys. Rev. E \textbf{82}, 041607 (2010)

\bibitem{Luis2017}

E.~E.~M.~Luis, T.~A.~de Assis and S.~C.~Ferreira, Phys. Rev. E \textbf{95}, 042801 (2017)

\bibitem{Luis2019}

E.~E.~M.~Luis, T.~A.~de Assis, S.~C.~Ferreira and R.~F.~S.~Andrade, Phys. Rev. E \textbf{99}, 022801 (2019)

\bibitem{Pereira2019}

A.~J.~Pereira, S.~G.~Alves and S.~C.~Ferreira, Phys. Rev. E \textbf{99}, 042802 (2019)

\bibitem{Chatraphorn2001}

P.~P.~Chatraphorn, Z.~Toroczkai and S.~Das Sarma, Phys. Rev. B \textbf{64}, 205407 (2001)

\bibitem{Chatraphorn2002}

P.~P.~Chatraphorn and S.~Das Sarma, Phys. Rev. E \textbf{66}, 041601 (2002).

\bibitem{Mal2010}

B.~Mal, S.~Ray, and J.~Shamanna, Eur. Phys. J. B \textbf{82}, 341-347 (2011).

\bibitem{Mal2017}

B.~Mal, S.~Ray, and J.~Shamanna, AIP Conference Proceedings \textbf{1832}, 040021 (2017).

\bibitem{Reis2010}

F.~D.~A.~Aarão Reis, Phys. Rev. E, \textbf{81}, 041605 (2010).

\bibitem{To2018}

T.~B.~T.~To, V.~B.~Sousa, F.~D.~A.~Aarão Reis, Physica A \textbf{511}, 240-250 (2018).

\bibitem{Lai1991}

Z.~W.~ Lai and S.~Das Sarma, Phys. Rev. Lett. \textbf{66}, 2348 (1991).

\bibitem{Herring1950}

C.~Herring, J. Appl. Phys. \textbf{21}, 301 (1950).

\bibitem{Mullins1957}

W.~W.~Mullins, J. Appl. Phys. \textbf{28}, 333 (1957).

\bibitem{Bartelt1992}

M.~C.~Bartelt and J.~W.~Evans, Phys. Rev. B \textbf{46}, 12675 (1992)

\bibitem{Amar2004}

J.~G.~Amar and M.~N.~Popescu, Phys. Rev. B \textbf{69}, 033401 (2004)

\bibitem{Maass2012}

M.~K\"orner, M.~Einax, and P.~Maass, Phys. Rev. B \textbf{86}, 085403 (2012)

\bibitem{Kim1989}

J.~M.~Kim and J.~M.~Kosterlitz, Phys. Rev. Lett. \textbf{62}, 2289 (1989).

\bibitem{Wolf1989}

D.~E.~Wolf and J.~Kertész, Phys. Rev. Lett. \textbf{63}, 1191 (1989).

\bibitem{Punyindu1998}

P.~Punyindu Chatraphorn and S.~Das Sarma, Phys. Rev. E \textbf{57}, 4863(R) (1998).

\bibitem{Punyindu2002}

P.~Punyindu Chatraphorn and S.~Das Sarma, Phys. Rev. E \textbf{66}, 041601 (2002).

\bibitem{Punyindu2001}

P.~Punyindu Chatraphorn Z.~Toroczkai, and S.~Das Sarma, Phys. Rev. B \textbf{64}, 205407 (2001).

\bibitem{Family_Vicsek_1985}

F.~Family and T.~Vicsek, J. Phys. A \textbf{18}, L75-L81 (1985).

\bibitem{Richards1991}

P.~M.~Richards, Phys. Rev. B \textbf{43}, 6750 (1991).

\bibitem{Chame2004}

A.~Chame and F.~D.~A.~Aarão Reis, Surf. Sci. \textbf{553}, 145-154 (2004).

\bibitem{Kardar1986}
	
M.~Kardar, G.~Parisi, and Y.~C.~Zhang, Phys. Rev. Lett. \textbf{56}, 889 (1986).
	
\bibitem{Takeuchi2011}

K.~A.~Takeuchi, M.~Sano, T.~Sasamoto, and H.~Spohn (2011), Sci. Rep. \textbf{1}, 34 (2011).

\bibitem{Bertini1997}

L.~Bertini and G.~Giacomin, Comm. in Math. Phys., \textbf{183}, 571-607 (1997).

\bibitem{Krug1988}

J.~Krug and H.~Spohn, Phys. Rev. A \textbf{38}, 4271 (1988).

\bibitem{Krug1993}

J.~Krug, M.~Plischke, and M.~Siegert, Phys. Rev. Lett. \textbf{70}, 3271 (1993).

\bibitem{Ehrlich1966}
	
G.~Ehrlich and F.~G.~Hudda, J. Chem. Phys. \textbf{44}, 1039 (1966).
	
\bibitem{Schwoebel1966}

R.~L.~Schwoebel and E.~J.~Shipsey, Journal of Applied Physics \textbf{37}, 3682 (1966).
	
\bibitem{Schwoebel1969}

R.~L.~Schwoebel, Journal of Applied Physics \textbf{40}, 614 (1969).

\bibitem{Amar_Family_95}

J.~G.~Amar and F.~Family, Phys. Rev. Lett. \textbf{74}, 11 (1995).

\bibitem{Amar2001}

J.~G.~Amar, M.~N.~Popescu and F.~Family, Surf. Sci. \textbf{491}, 239 (2001)

\bibitem{Reis2004}

F.~D.~A.~Aarão Reis, Phys. Rev. E \textbf{70}, 031607 (2004).

\bibitem{Neave1983}

J.~H.~Neave, B.~A.~Joyce, P.~J.~Dobson, and N.~Norton, Appl. Phys. A \textbf{31}, 1-8 (1983).

\bibitem{Alexandre1985}

F.~Alexandre, L.~Goldstein, G.~Leroux, M.~C.~Joncour, H.~Thibierge, and E.~V.~K.~Rao, Journal of Vacuum Science and Technology B \textbf{3}, 950 (1985).

\bibitem{Kunkel1990}

R.~Kunkel, B.~Poelsema, L.~K.~Verheij, and G.~Comsa, Phys. Rev. Lett. \textbf{65}, 733 (1990).

\bibitem{Smilauer1993}

P.~$\check{\text{S}}$milauer, M.~R.~Wilby, and D.~D.~Vvedensky, Phys. Rev. B \textbf{47}, 4119 (1993).

\bibitem{Assis2013}

T.~A.~de Assis and F.~D.~A.~Aarão Reis, J. Stat. Mech. P10008 (2013).

\bibitem{Assis2015}

T.~A.~de Assis and F.~D.~A.~Aarão Reis, J. Stat. Mech. P06023 (2015).

\bibitem{Burioni1995}

R.~Burioni, Phys. Rev. E \textbf{51}, 5426 (1995).

\bibitem{Park2001}

S.~C.~Park, J.~M.~Park, and D.~Kim, Phys. Rev. E \textbf{65}, 015102(R) (2001).

\end{thebibliography}

{}

\end{document}